# Lasing Action with Gold Nanorod Hyperbolic Metamaterials

*Rohith Chandrasekar[1], Zhuoxian Wang[1], Xiangeng Meng[1,*], Mikhail Y. Shalaginov[1], Alexei Lagutchev[1], Young L. Kim[2], Alexander Wei[3], Alexander V. Kildishev[1], Alexandra Boltasseva[1], and Vladimir M. Shalaev[1,*]*

[1]School of Electrical and Computer Engineering and Birck Nanotechnology Center, Purdue University, West Lafayette, IN 47907, USA

[2]Weldon School of Biomedical Engineering, Purdue University, West Lafayette, IN 47907, USA

[3]Department of Chemistry, Purdue University, West Lafayette, IN 47907, USA

## ABSTRACT

Coherent nanoscale photon sources are of paramount importance to achieving all-optical communication. Several nanolasers smaller than the diffraction limit have been theoretically proposed and experimentally demonstrated using plasmonic cavities to confine optical fields. Such compact cavities exhibit large Purcell factors, thereby enhancing spontaneous emission, which feeds into the lasing mode. However, most plasmonic nanolasers reported so far have employed resonant nanostructures and therefore had the lasing restricted to the proximity of the resonance wavelength. Here, we report on an approach based on gold nanorod hyperbolic metamaterials for lasing. Hyperbolic metamaterials provide broadband Purcell enhancement due to large photonic density of optical states, while also supporting surface plasmon modes to deliver optical feedback for lasing due to nonlocal effects in nanorod media. We experimentally demonstrate the advantage of hyperbolic metamaterials in achieving lasing action by its comparison with that obtained in a metamaterial with elliptic dispersion. The conclusions from the experimental results are supported with numerical simulations comparing the Purcell factors and surface plasmon modes for the metamaterials with different dispersions. We show that although the metamaterials of both types support lasing, emission with hyperbolic samples is about twice as strong with 35% lower threshold vs. the elliptic ones. Hence, hyperbolic metamaterials can serve as a convenient platform of choice for nanoscale coherent photon sources in a broad wavelength range.

TOC Graphic

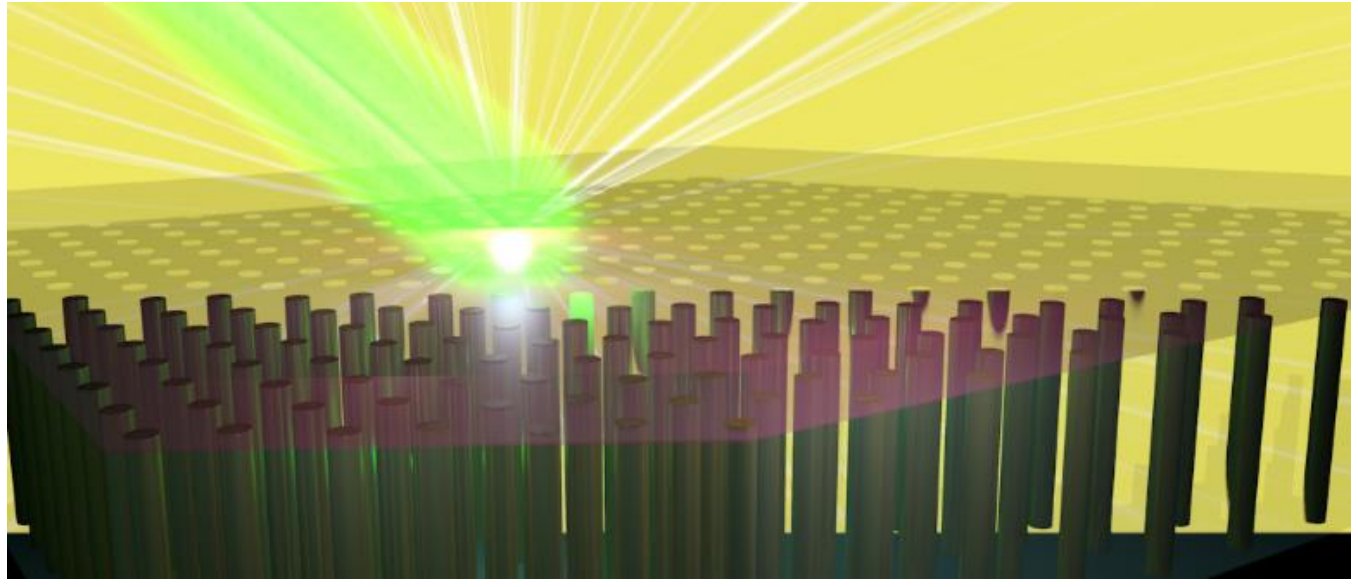

We study lasing in two gold nanorod arrays coated with Rhodamine 101, one exhibiting hyperbolic dispersion at the lasing wavelength, the other with elliptic dispersion. Experiments show the hyperbolic metamaterial provides stronger emission with reduced threshold.

Since its invention in 1960[1], the laser has seen tremendous developments and has quickly revolutionized fundamental and applied fields such as metrology, medicine, data storage, fabrication and telecommunications among others[2]. With the ever growing need for data transfer speeds and compact devices, several efforts have been made in miniaturizing the laser for on-chip integration[3-7]. While photonic cavities have proven to exhibit high-Q factors enabling strong lasing, their miniaturization to the nanoscale is not viable since the diffraction limit requires the cavity length to be at least half the lasing wavelength[8-10]. In contrast, plasmonic cavities, which can be employed to achieve optical amplification and lasing action with charge oscillations, have successfully led to the design of coherent photon sources no longer limited by the diffraction limit[10]. Coupling of emitters with the strongly confined electromagnetic fields associated with plasmonic oscillations can significantly enhance spontaneous emission in certain modes, a process known as the Purcell effect[11]. This effect yields a redistribution of spontaneous emission over wavevector space (k-space), with light preferentially coupled to the lasing mode and inhibited in other modes[12].

Several plasmonic lasers have already been demonstrated using various geometries, such as semiconductor-dielectric-metal hybrid cavities[7], metal-insulator-metal waveguides[13], whispering gallery cavities[14], core-shell particles[15, 16], nanohole/particle arrays[17-19]. These methods generally used geometries with strong cavity resonances to gain Purcell enhancement and hence lasing as described above. The Purcell enhancement arising from the resonance of metallic structures usually exhibits a relatively narrow bandwidth, thus restraining the frequency of lasing to be achieved. An alternative approach to gain Purcell enhancement is based on non-resonant structures, which can be achieved by engineering the dispersion of metamaterials[20-22]. In this work, we report on the use of nanorod-based metamaterial with hyperbolic dispersion, namely hyperbolic metamaterials (HMMs) to achieve efficient lasing. Our nanorod-based metamaterials are composed of vertically aligned gold nanorods coated with a polyvinyl alcohol (PVA) film embedded with Rhodamine 101 (R101) dye molecules. The dispersion of the metamaterial can be controlled by changing the fill ratio of the metal[23]. Our experimental results show that HMM exhibits larger enhancement of lasing over the elliptic metamaterial (EMM), which could be related to both Purcell enhancement and field coupling to surface plasmon modes due to nonlocal effects in the nanorod structure. Our simulations show that HMM gives a larger magnitude of Purcell enhancement than EMM, which could feed a higher portion of spontaneous emission into the lasing mode. Concurrently, simulations indicate that nonlocal effects give rise to a more profound resonance in HMM, which could favor the formation of optical feedback for lasing. Due to both of these effects, nanorod-based hyperbolic metamaterials could serve as an efficient platform for lasing. Furthermore, the nanorod-based metamaterials are versatile to be integrated with a broad range of optical gain media to achieve lasing at a desired frequency.

HMMs are a special class of highly anisotropic metamaterials with effective dielectric permittivities of opposite signs for orthogonal tensor components[24-28], as shown in Equation 1. Longitudinal and transverse effective permittivities, $\varepsilon_{\perp}$ and $\varepsilon_{\parallel}$, can be calculated using the Maxwell-Garnett effective medium theory[29], as shown in Equations 2 and 3.

$$\varepsilon = \begin{pmatrix} \varepsilon_{\parallel} & 0 & 0 \\ 0 & \varepsilon_{\parallel} & 0 \\ 0 & 0 & \varepsilon_{\perp} \end{pmatrix} \tag{1}$$

$$\varepsilon_{\parallel} = \varepsilon_d \left[\frac{\varepsilon_m(1+f)+\varepsilon_d(1-f)}{\varepsilon_m(1-f)+\varepsilon_d(1+f)}\right] \tag{2}$$

$$\varepsilon_{\perp} = f\varepsilon_m + (1-f)\varepsilon_d \tag{3}$$

So far two types of HMMs have been classified. Type-I HMMs are typically based on aligned nanowire/nanorod arrays usually having $\varepsilon_{\perp} < 0$ and $\varepsilon_{\parallel} > 0$[23, 30-37], whereas type-II HMMs are based on metal-dielectric stacks usually having $\varepsilon_{\perp} > 0$ and $\varepsilon_{\parallel} < 0$[27, 38-40]. By adjusting the chemical composition of the metamaterial or the metal fill ratio, the dispersion of the metamaterial can be tuned from elliptic to hyperbolic. It has been shown that metamaterials exhibiting hyperbolic properties can support unique optical waves with very small or large mode indices, allowing for stronger light-matter interaction[32, 33]. Furthermore, the fields in the nanorod metamaterial can exhibit strong spatial variation yielding a nonlocal response[32, 41].

Several physical phenomena have already been demonstrated either theoretically[42-44] or experimentally[21, 22, 45-52] with HMMs, such as focusing[42], sub-diffraction imaging[45], thermal emission enhancement[43, 46] and spontaneous emission enhancement[21, 22, 44, 47-52]. HMMs exhibit large Purcell factors due to the singularity in their local density of optical states (LDOS), a property that manifests due to hyperbolic dispersion[20, 48]. Radiative decay rate enhancement of emitters such as dyes and nitrogen-vacancy centers using HMMs has been studied extensively[22, 23, 47-53], with as high as 76-fold enhancement of spontaneous emission rate being reported[22]. On the other hand, it has been shown that the Purcell effect could contribute to efficient optical amplification and lasing action, provided that a portion of the enhanced spontaneous emission feeds into the lasing mode[12]. To our knowledge, so far there is only one report that addressed the possibility to achieve stimulated emission from a HMM comprised of Ag and $MgF_2$ layer stacks[53]. Although a reduced threshold was observed in a HMM when compared to a reference device based on a bare Ag film, the emission efficiency from the HMM was obviously lower than the latter. Thus, the full potential of HMM to achieve lasing with low threshold and high efficiency needs to be explored further.

In this work, two gold nanorod arrays embedded in anodic alumina templates have been fabricated, exhibiting hyperbolic (labeled HMM) and elliptic dispersion (labeled EMM) at the emission wavelength of R101 ($\lambda$ = 606 nm), according to local effective medium theory (EMT, see Eqns 1-3). The different dispersion characteristics are achieved by altering the metal fill ratio using different nanorod diameters (Supporting Information S1).

Arrays of gold nanorods were fabricated by electrodeposition within nanoporous aluminum oxide membranes that were prepared by the anodization of Al films deposited on tantalum pentoxide and Au-coated glass substrates in 0.3 M $H_2SO_4$ (Supporting Information S1). Anodization was performed at 30 V for the HMM and at 25 V for the EMM, to yield approximate pore diameters of 40±2 and 25±2.5 nm respectively, surface densities of 35% and 14% respectively, and nanopore heights of 250 nm. Gold nanorods were electrodeposited according to a procedure as previously described under galvanostatic conditions[54, 55] using a current density of 0.5 $mA/cm^2$, with constant electrodeposition up to the maximum height allowed by the nanoporous alumina templates, at which point a distinct drop in voltage (>20%) was observed. SEM images of the nanorod-based HMM and EMM indicate that the nanorods have uniform diameters and are well dispersed within the $Al_2O_3$ matrix, which has an approximate pore-to-pore distance of 60 nm for each sample (Figure 1b,c).

Transmission spectra using TM-polarized light at different angles of incidence show that the HMM sample produces complete extinction up to 550 nm, whereas the EMM sample exhibits transmission minima at 530 and 710 nm (Figure 1d,e). These minima can be assigned to the epsilon-near-pole (ENP, $\varepsilon_{\parallel} \to \infty$) and epsilon-near-zero (ENZ, $\varepsilon_{\perp} = 0$) resonances[56]. As follows from Equation 2, the ENP resonance is purely dependent on the permittivities of the metal and dielectric, and hence does not vary with the metal fill ratio. However, the ENZ response (and subsequent cutoff wavelength for hyperbolic dispersion) can be tuned easily by adjusting the fill ratio. In the case of HMM, only one resonance band is observed since the ENP and ENZ responses overlap starting at λ = 530 nm, whereas EMM has distinct ENP (λ = 530 nm) and ENZ (λ = 710 nm) responses, as seen in Figure 2a,b. Simulations of the optical responses of the HMM and EMM structures also match the experimental extinction data (Supporting Information S2); effective Maxwell-Garnett permittivities were calculated based on the fill ratios for each sample and the permittivity values of bulk Au and amorphous $Al_2O_3$ (Figure 2c,d). Both the extinction and permittivity plots confirm that the HMM exhibits hyperbolic dispersion and the EMM exhibits elliptic dispersion at the same wavelength as the central emission of R101. In addition, we have calculated the iso-frequency curves for both HMM and EMM based on the effective medium permittivities (Supporting Information S3).

The Purcell factor, $F_P$, is defined as the ratio of the total decay rate from a dipole source situated in an enhancing medium, such as a cavity, to that in the superstrate (PVA in this case). HMMs, due to their singularity in LDOS, offer a non-resonant method of achieving large Purcell factors for emitters[48]. We estimated the Purcell factors for our HMM and EMM structures by simulating the coupling of dipole emitters with our metamaterials, defined by local EMT. Using the standard Green's function formalism[57], we calculated the Purcell factors for a dipole embedded in PVA (refractive index n = 1.5) placed 20 nm above HMM and EMM supported on glass substrates (n = 1.5), as shown in Figure 3a and 3b. In the calculation, the polarization of the dipole is addressed as well. The calculation results show that the HMM provides a Purcell factor of 5.75 for a dipole parallel to the HMM surface, and 12.21 for a dipole perpendicular to the surface at the central emission wavelength of R101 (λ = 606 nm). On the other hand, the EMM provides a Purcell factor of 1.5 and 2.15 for parallel and perpendicular orientations, respectively. On average the HMM provides an enhancement of 4.6 times over the EMM. The calculation also shows that the Purcell factor strongly depends on the distance of the dipole from the metamaterial surface (Figure 3c). For dipoles very close to the metamaterial surface (~5 nm), the Purcell factor for the HMM is extremely large, reaching up to 400, while the EMM only provides an enhancement of 46. As the dipole is moved away from the surface, the Purcell factor for the EMM decays much quicker than that for HMM, reaching ~1 within a 40 nm distance, while it is still slightly larger than 1 for the HMM at a distance of 100 nm.

In order to further study the decay channels providing this Purcell enhancement, we have also calculated the inherent plasmonic modes (often named as high-k modes) that can be excited in our HMM and EMM structures. Figure 4 shows the *k*-space dissipated power density, $\log_{10}\left[k_0 dF_P^{ave} / dk_{||}\right]$ (Supporting Information S4), calculated for a dipole with the averaged orientation, located 20 nm above the surface of HMM and EMM, defined by local EMT. In both HMM and EMM, we can see the modal gap, i.e., wavelength range where very few plasmonic modes are allowed between ENP and ENZ resonances. We see that at 606 nm, EMM falls in this modal gap with almost no propagating modes, while HMM provides much more modes to enhance the emitter's decay rate, for even very large $k_{||}$ (in-plane wavevector). Due to this large number of allowed modes, we clearly see that our HMM provides significant Purcell enhancement over our EMM structure, and is therefore expected to enhance spontaneous emission greatly, which in turn could feed into lasing modes.

The nanorod arrays were coated with a ~2 μm thick film of PVA embedded with R101 dye (10 mM). A frequency-doubled Nd:YAG picosecond laser (λ = 532 nm, pulse width = 400 ps, repetition rate = 1 Hz) was used to pump the samples. The emission from the samples was collected with a fiber, which was fed to a spectrometer equipped with a charge-coupled device (Supporting

Information S5). Figure 5a,b shows the evolution of the emission spectra with the pump energy for both HMM and EMM. Obvious spectral narrowing was observed for both HMM and EMM when the pump energy was increased up to ~3.2 μJ and 4.0 μJ respectively. The full widths at half maximum (FWHM) of the emission spectra from HMM and EMM samples were ~5.8 nm and 6.1 nm, respectively, when the samples were pumped at ~5.5 μJ (Supporting Information S6). The emission linewidth for both HMM and EMM was found to be similar to those demonstrated in many other reports using laser dyes as gain media[15, 58]. At a pump energy of ~5.5 μJ, the emission intensity from HMM was also twice as strong as that from EMM. From the plot of pump-dependent peak emission intensity (Figure 5c,d), we can see a clear threshold for the two systems, with HMM having a lower threshold (~3.2 μJ) than EMM (4 μJ). We have examined the threshold behavior at 5 points on each sample and found that the HMM achieves an average of 35% reduction in threshold energy compared to EMM (Supporting Information S7).

As control samples, we have studied the emission from a bare glass substrate and a 250-nm thick gold film, both coated with R101 in PVA (Supporting Information S8). Both samples only show spontaneous emission, without any spectral narrowing. We have also fabricated a lamellar HMM with ten alternating layers of Au (8.9 nm) and $Al_2O_3$ (16.1 nm), which has the same metal fill ratio as the nanorod-based HMM. We applied the gain medium and conducted lasing measurements in the same way as described above and enhanced emission was collected from the sample based on the lamellar HMM; however, no lasing was observed before the sample was damaged by the pump pulse (Supporting Information S9). Therefore, the nanorod-based HMM provides a significant enhancement over its lamellar counterpart, exhibiting low-threshold lasing action.

As described above, the HMM exhibits a higher Purcell factor at the lasing wavelength compared to EMM, which could partially account for low-threshold behavior observed in HMM due to spontaneous emission being fed into lasing mode. To further understand the lasing mechanism, we have computed optical responses and field distributions of the nanorod metamaterials (Supporting Information S10). In comparison to EMM, HMM exhibits a more distinguishable resonance arising from nonlocal effects[32], as seen in the optical absorption spectrum (Fig. S10c,e), and a higher magnitude of averaged local fields (Fig. S10d,f) at ~606 nm. Along with a higher Purcell factor, the stronger resonance and local fields exhibited in HMM could provide stronger feedback for lasing and thus lead to a lower threshold. Our further calculation based on realistic gain used in this work shows that the optical absorption becomes negative at ~606 nm, indicating that the losses of the system are completely compensated by the gain (Supporting Information S11). Bare metal film and lamellar HMM also provide Purcell enhancement to some extent, but lack the resonant behavior needed to serve as feedback for lasing action.

It is noteworthy to mention that the lasing herein is distinct from that reported in a previous work based on tilted silver nanorods in terms of sample configuration and lasing mode dispersion[59]. In Ref. 59, the tilted nanorods, acting as random scatterers, have relatively large distributions in both length and diameter. The dyes are infiltrated into the random nanorod arrays where significant scattering is naturally anticipated. In contrast, in this work the dyes are accumulated atop a relatively smooth metamaterial surface so light scattering can be neglected. In addition, the features of lasing spectra in Ref. 59 are consistent with the behavior exhibited in random lasers with coherent feedback[60], and thus are distinct from the observations in this work.

In conclusion, we have demonstrated lasing action in metamaterials based on gold nanorod arrays coated with thin films of PVA embedded with R101, which can be tuned to exhibit hyperbolic or elliptic dispersion based on their metal fill ratio. Both metamaterials support lasing action, with emission from the HMM being twice as strong as that from the EMM while also exhibiting 35% lower threshold. One interesting direction for future study is to infuse dyes directly into the matrix of the nanorod-based HMM which would greatly enhance the emitter-field coupling and consequently give much lower thresholds for lasing. The HMM achieves Purcell enhancement via non-resonant means, suggesting its application as a source for coherent photon emission and Forster energy transfer[61] in a broadband wavelength range.

ASSOCIATED CONTENT

**Supporting Information Available:** Fabrication and characterization details, simulated extinction spectra of nanorod-based metamaterials, iso-frequency curves for metamaterials, equations for calculating the k-space dissipated power density, spectra narrowing and threshold distribution for nanorod HMM and EMM, emission from glass, gold film, and lamellar hyperbolic metamaterial sample, local field distributions, and loss compensation of plasmonic resonance in HMM.

AUTHOR INFORMATION

**Corresponding Author**

mengxiangeng@gmail.com (X.M.); shalaev@purdue.edu (V.M.S)

**Author Contributions**

R.C., X.M. and V.M.S designed this work. R.C. fabricated the nanorod samples and made characterizations, and Z.W. fabricated the lamellar HMM sample. R.C., Z.W., and X.M. conducted the lasing experiments. A.W. and A.L. assisted the nanorod fabrication and Y.L.K assisted lasing experiments. R.C. conducted simulations of extinction spectra, X. M. and M.Y.S. conducted simulations of Purcell enhancement and study of nonlocal effects and study of nonlocal effects. A.V.K. conducted simulations of field distributions and loss compensation in nanorod metamaterials. R.C. and Z.W. contributed equally to this work. R.C. wrote the paper and all the other authors commented on the manuscript. All authors have given approval to the final version of this manuscript.

**Funding Sources**

Authors acknowledge support from the Office of Naval Research grant #N00014-13-1-0649.

## ACKNOWLEDGMENT

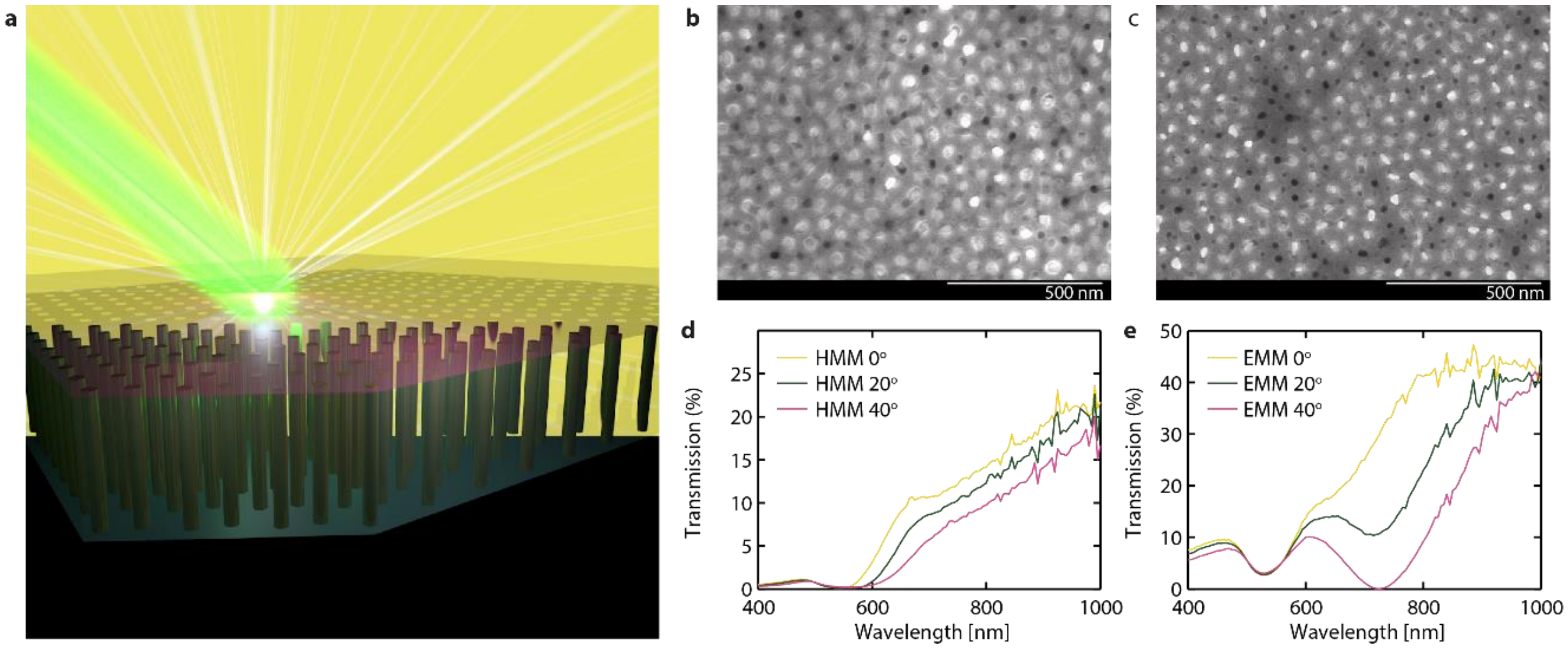

Figure 1. (a) Scheme of lasing action from nanorod-based metamaterials. (b,c) Top-view SEM images of HMM and EMM, with metal fill ratios of 35% and 14%, respectively. (d,e) Transmission spectra of HMM and EMM illuminated with TM–polarized light at 0°, 20° and 40° incidence.

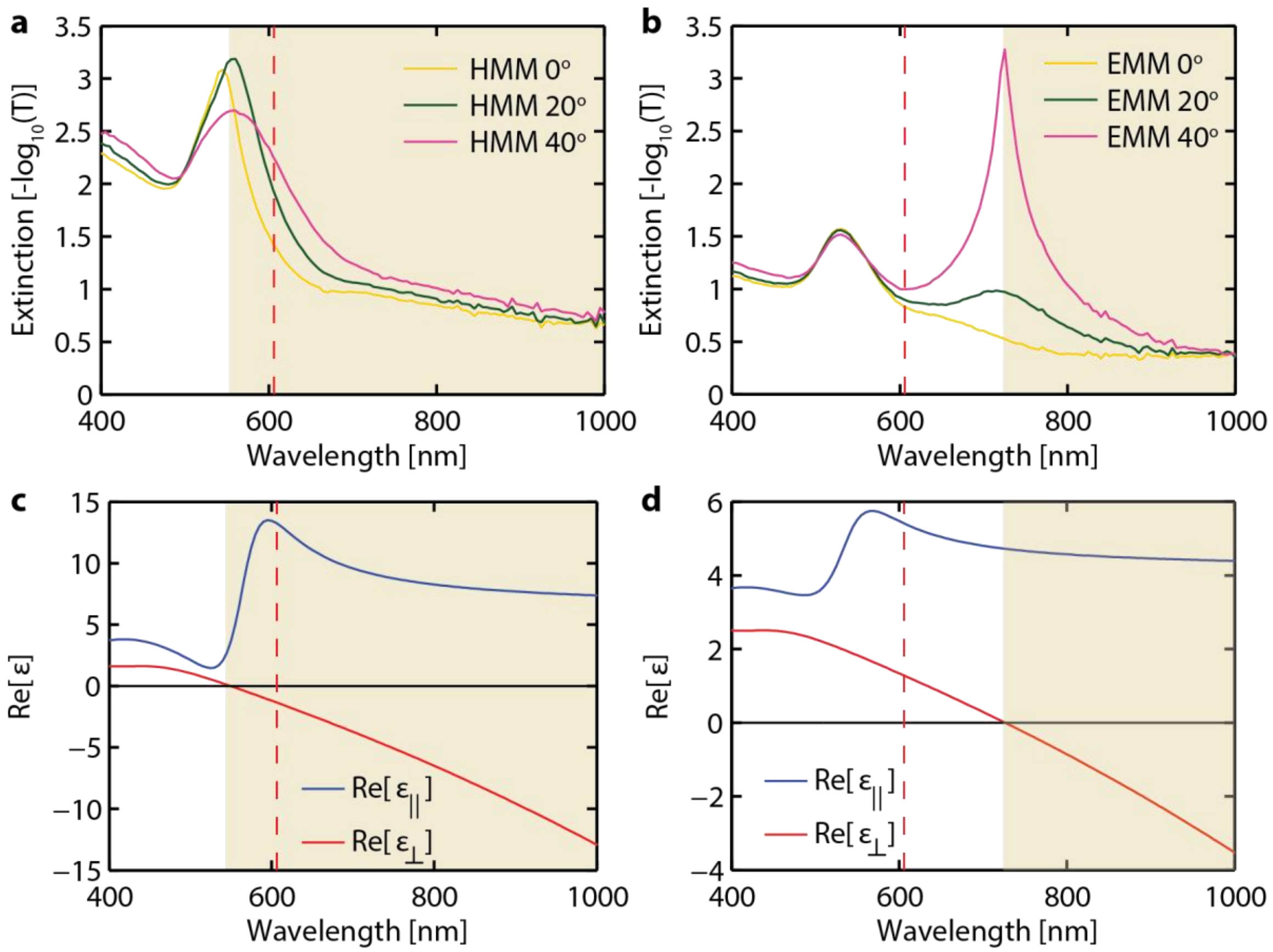

Figure 2. (a,b) Experimental extinction spectra for HMM (a) and EMM (b). (c,d) Effective anisotropic permittivities of HMM (c) and EMM (d) estimated using Maxwell-Garnett theory. Shaded regions indicate wavelength range where the metamaterial exhibits hyperbolic Type-I dispersion. Dashed red line indicates the central emission wavelength of R101 dye (606 nm).

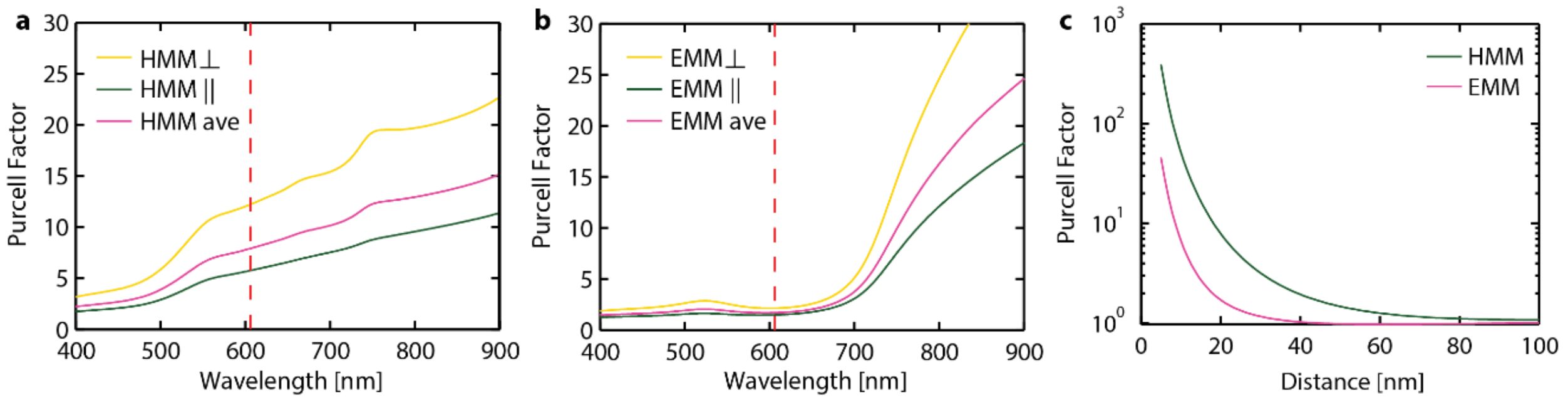

Figure 3. Theoretical estimations of Purcell factor versus emission wavelength for a dipole located 20 nm above HMM (a) and EMM (b), as well as dependence of Purcell factor on the distance of dipole from metamaterial surface (c) at the central emission wavelength of R101 (606 nm, indicated with a red dashed line in (a) and (b)). Yellow, green, and pink curves in (a) and (b) refer to dipoles oriented perpendicular, parallel to the metamaterial surface, and averaged, respectively. The Purcell factor in (c) is plotted for the averaged dipole orientation.

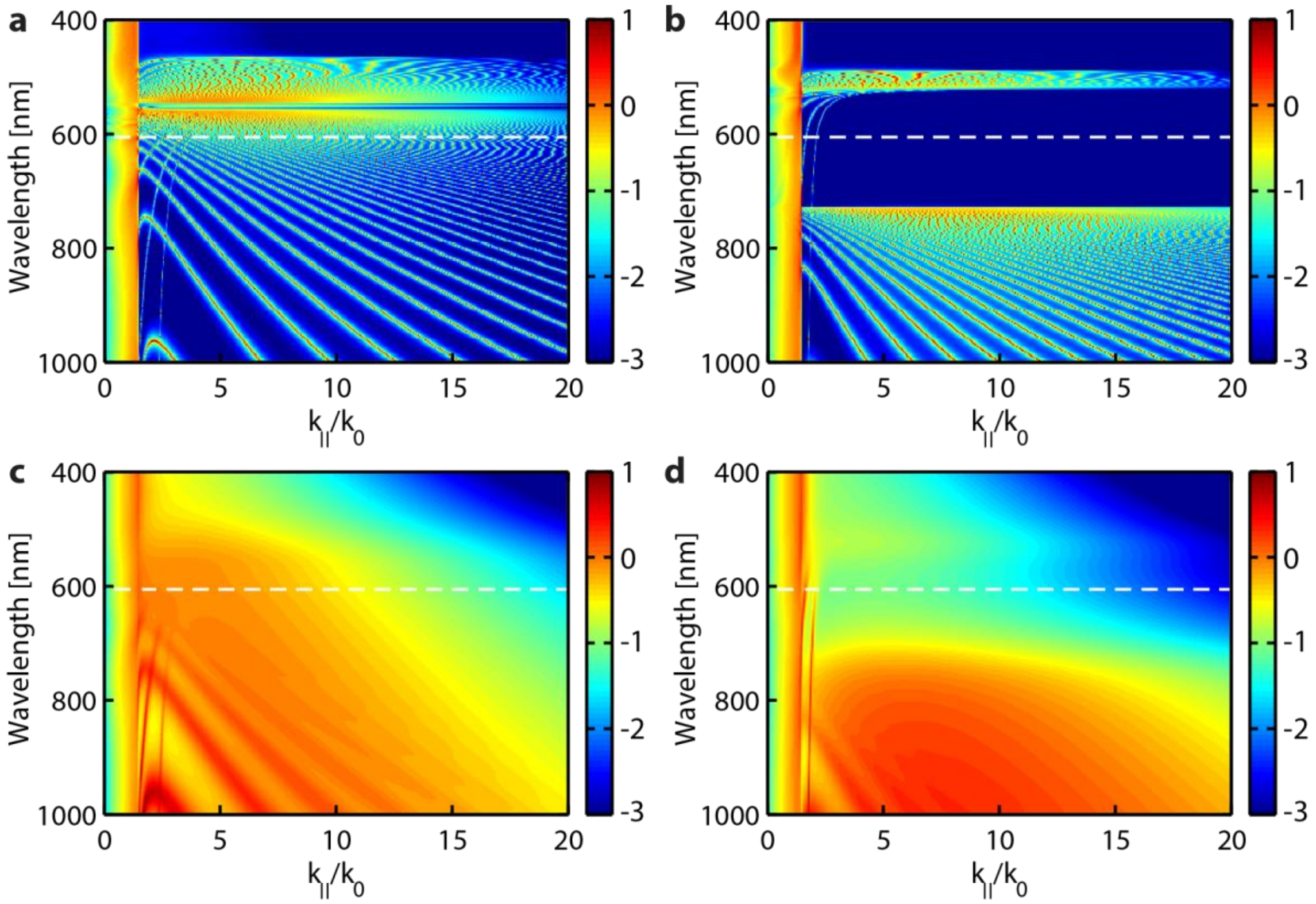

Figure 4. k-space dissipated power density for a dipole placed 20 nm above HMM (a) and EMM (b), without considering losses. (c) and (d) correspond to the cases of (a) and (b) with actual losses. At 606 nm (white dotted line), HMM provides many more inherent plasmonic modes than EMM. The white dashed lines indicate the position of the central emission wavelength of R101 (606 nm).

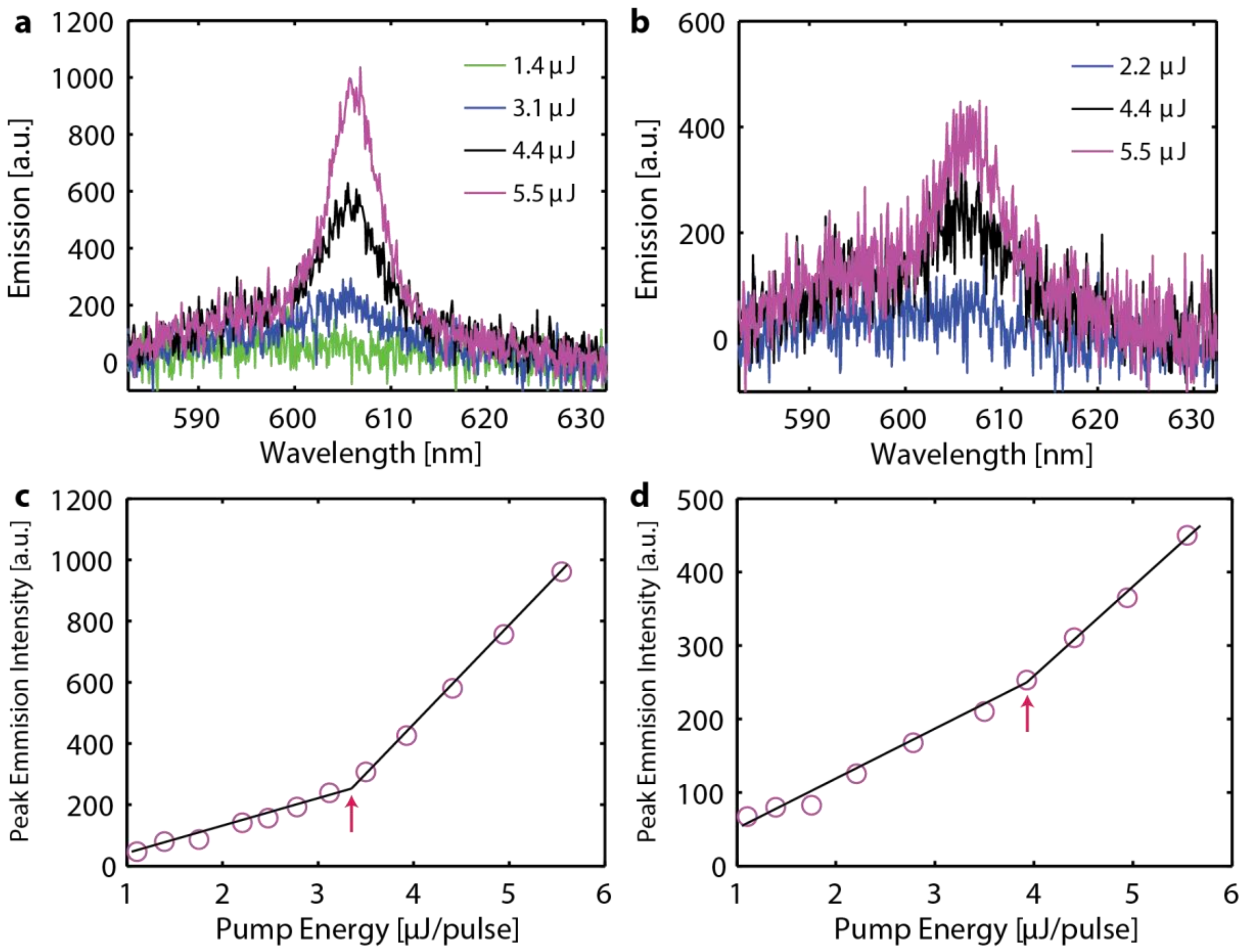

Figure 5. (a,b) Emission spectra recorded under various pump energies for HMM (a) and EMM (b). (c,d) Peak intensity plots for HMM and EMM; the slope kinks correspond to the lasing threshold energy (indicated by red arrows).

Supporting Information for

# Lasing Action with Gold Nanorod Hyperbolic Metamaterials

*Rohith Chandrasekar[1], Zhuoxian Wang[1], Xiangeng Meng[1,*], Mikhail Y. Shalaginov[1], Alexei Lagutchev[1], Young L. Kim[2], Alexander Wei[3], Alexander V. Kildishev[1], Alexandra Boltasseva[1], and Vladimir M. Shalaev[1,*]*

[1]School of Electrical and Computer Engineering and Birck Nanotechnology Center, Purdue University, West Lafayette, IN 47907, USA

[2]School of Biomedical Engineering, Purdue University, West Lafayette, IN 47907, USA

[3]Department of Chemistry, Purdue University, West Lafayette, IN 47907, USA

## S1. Fabrication and characterization

The arrays were fabricated by anodizing 250 nm-thick aluminum films deposited on glass substrates with 15 nm-thick tantalum pentoxide and 5 nm-thick gold under layers. All layers were deposited using an electron-beam evaporator (Leybold, Inc). The films were anodized in 0.3 M $H_2SO_4$ at 1°C until the aluminum film was fully anodized. Substrate anodization was performed at 30 V for the HMM cavity and at 25 V for the EMM cavity. A solution of 30 mM sodium hydroxide was used to remove the alumina barrier layer as well as widen the pores. For the HMM, the pores were widened for 60 s, while they were widened only for 25 s for the EMM. This fabrication procedure yielded HMM and EMM alumina templates with approximate pore diameters and surface densities of 40 nm and 35% and 25 nm and 14%, respectively (Figure S1).

Gold nanorods were grown to the full thickness of the template (~250 nm) by electrodeposition using Orotemp Gold Electrolyte (Technic Inc.) with a constant current of 0.5 $mA/cm^2$ using a Princeton Research Systems power supply, at which point a distinct drop in voltage (>20%) was observed. Samples were then cleaned using acetone, isopropanol and DI water and were dried under gentle nitrogen stream.

Transmission spectra were taken using a variable-angle spectroscopic ellipsometer (J.A. Woollam & Co.). SEM images are taken with a Hitachi Field-Emission Scanning Electron Microscope.

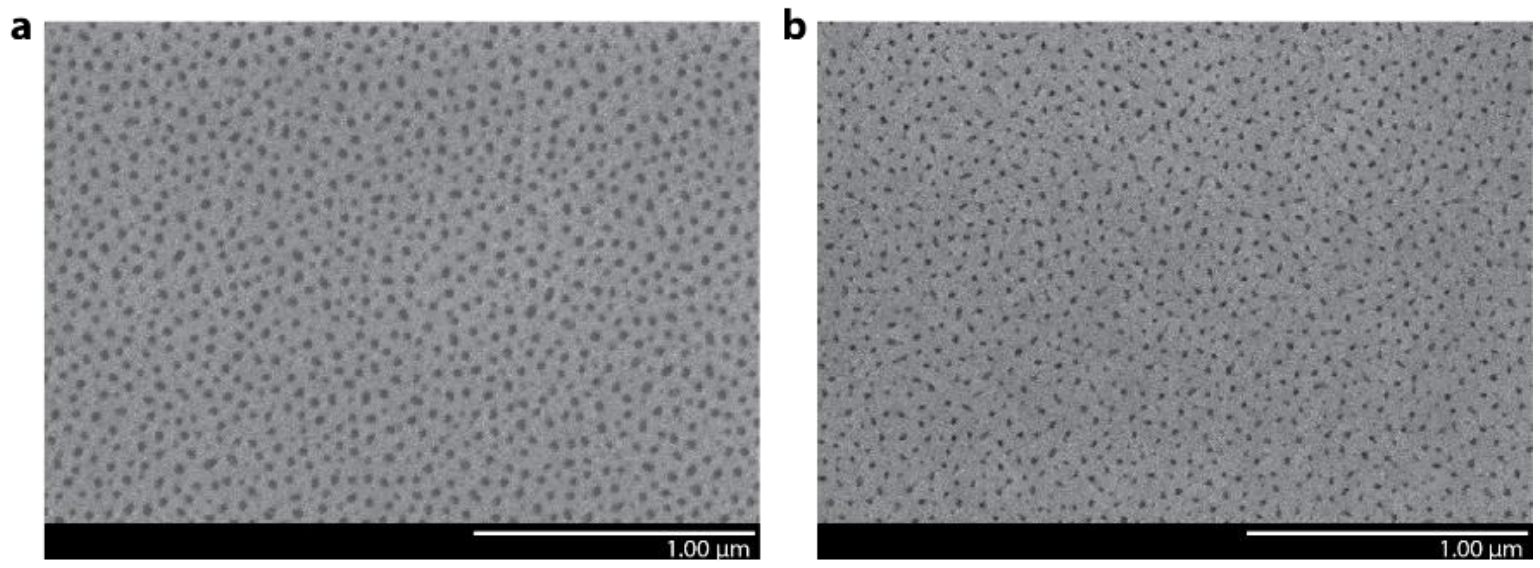

Figure S1. SEM images of alumina templates for (a) HMM and (b) EMM.

## S2. Simulations of nanorod metamaterials

We calculate the extinction spectra of the HMM and EMM for 0°, 20°, and 40° incidence using CST Microwave Studio. The simulated domain, unit cell as well as experimental and simulated extinction spectra are shown in Figure S2.

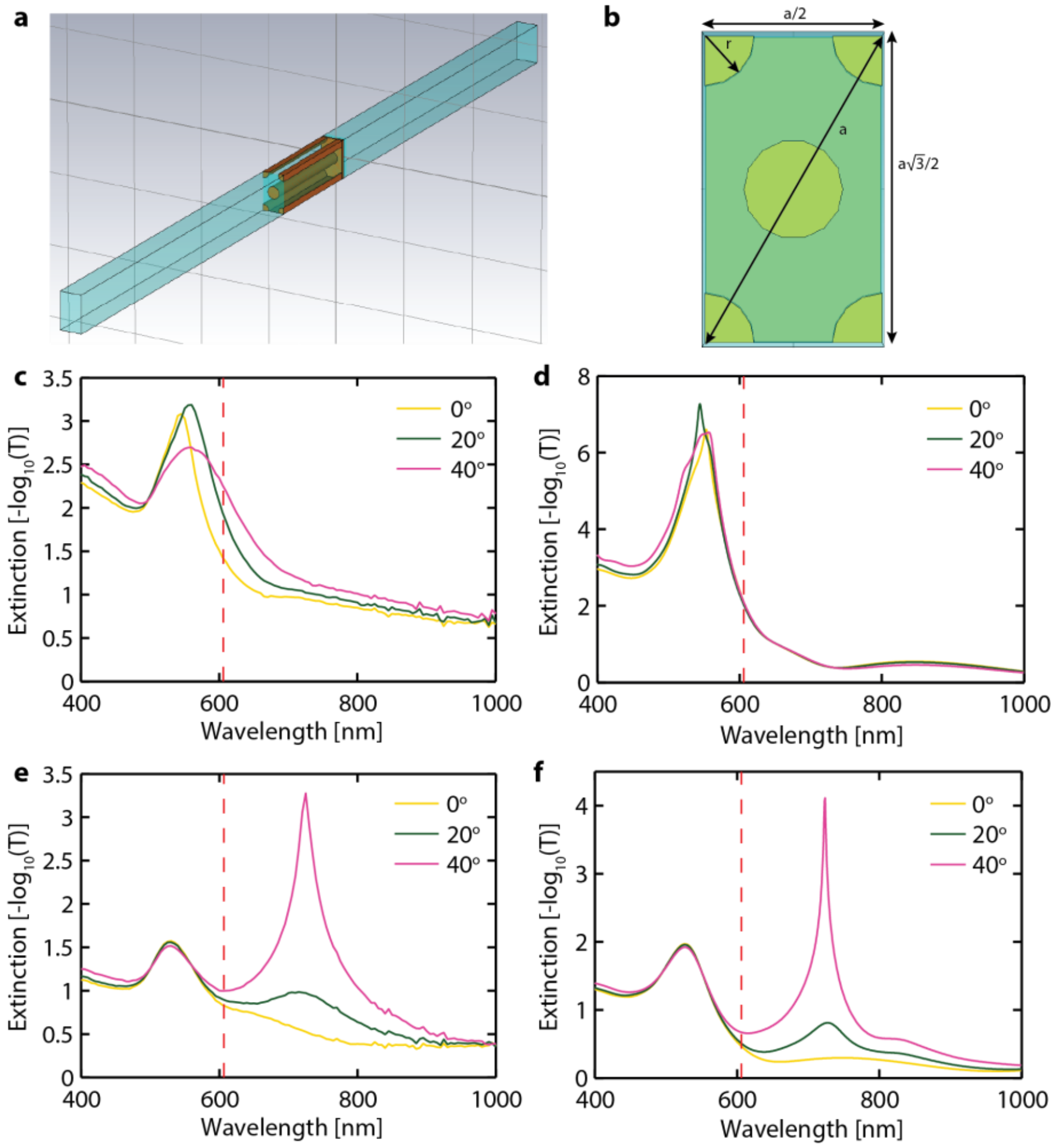

Figure S2. (a) Simulation domain, (b) unit cell of structure, (c,e) experimental extinction spectra of HMM (c) and EMM (e), simulated extinction spectra of HMM (d) and EMM (f). Red dashed line indicates the central emission wavelength from R101 (606 nm).

The structure, shown in Figure S2a, comprises of 1 μm glass substrate ($\varepsilon$ =2.25), covered with 15 nm tantalum pentoxide ($\varepsilon = 4.2$) and 5 nm gold layers. The gold nanorods have radius of $r$, and are 250 nm in length. The superstrate is air, and is 1 μm in length. The unit cell of the structure, shown in Figure S2b, has a diagonal length of $a$, and hence top width of $a/2$ and side length of $a\sqrt{3}/2$. For both EMM and HMM, the diagonal length $a$ is 120 nm. For the EMM, the rod diameter is 23.5 nm, yielding a fill ratio of ~14%. The HMM has rods with diameter 38 nm, which gives a fill ratio of ~35%. The alumina host has a refractive index of 1.77. The dispersion of gold is defined by a Drude term plus 2 critical points fit to Johnson-Christy data, with a loss factor of 2 [1]. The experimental extinction spectra for the HMM and EMM are shown in Figure S2c,e and the simulated fits are shown in Figures S2d,f. The simulated plots show the same spectral features as the experimental curves. While the simulated curves certainly show much sharper resonances, this discrepancy can be due to the roughness of the nanorods and alumina membrane, as well as inhomogeneity of the nanorods, which would cause inhomogeneous broadening.

## S3. Iso-frequency curves of nanorod metamaterials

We have calculated the iso-frequency curves (Equation 1 below) for 250-nm-thick nanorod metamaterial slabs with 35% (HMM) and 14% (EMM) metal fill ratios, as shown in Figure S3. The nanorod metamaterials have been simulated using local effective medium theory. $k_{||}$ is the in-plane wavevector, and $k_{\perp,\mathrm{eff}}$ is the effective wavevector of the propagating modes in the nanorod metamaterial.

$$\frac{k_{\perp,\mathrm{eff}}}{k_0} = \pm\sqrt{\varepsilon_{||}\left(1-\frac{\left(k_{||}/k_0\right)^2}{\varepsilon_{\perp}}\right)}, \quad (1)$$

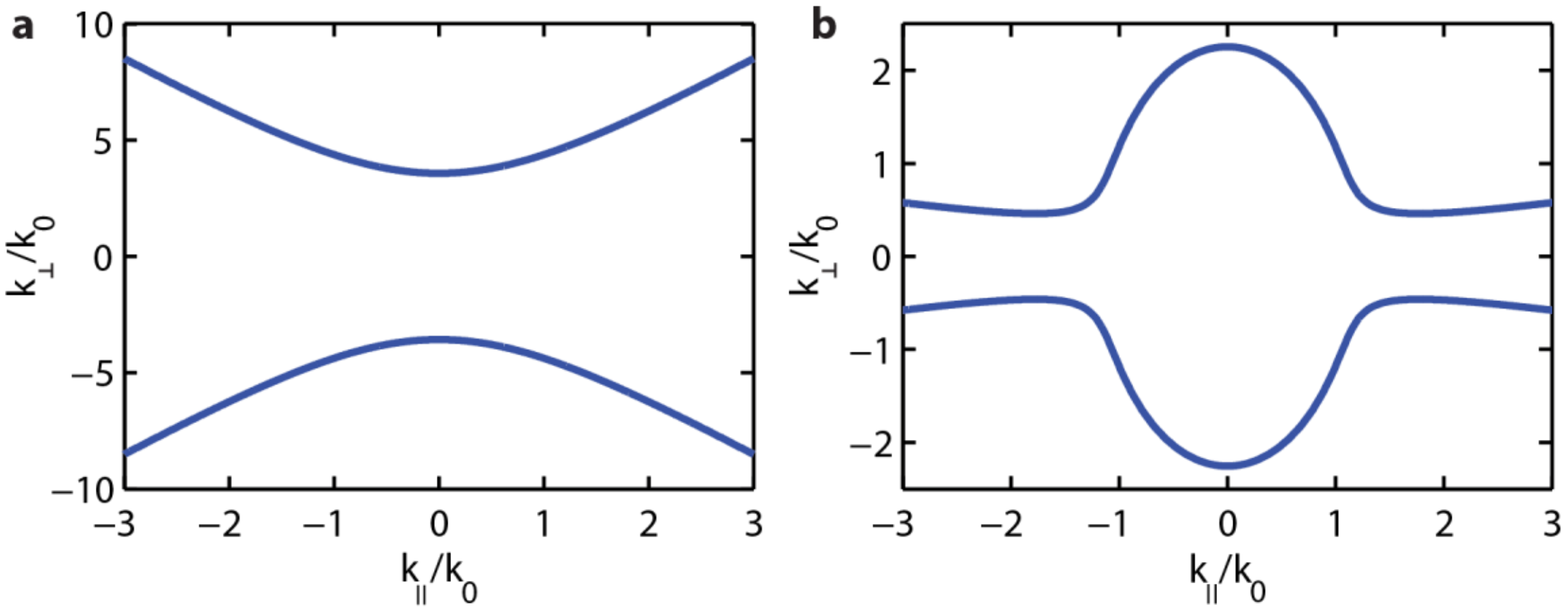

Figure S3. Iso-frequency curves at 606 nm for HMM (a) and EMM (b).

## S4. Simulations of plasmonic modes in nanorod metamaterials

The formulas used for the calculation of the k-space dissipated power density are [3]

$$\frac{dF_P^{\perp}}{ds} = \frac{3}{2}\mathrm{Re}\left\{ \frac{s^3}{s_{\perp,\mathrm{sup}}(s)\varepsilon_{\mathrm{sup}}^{\ 3/2}} \ 1 + \tilde{r}^{\mathrm{p}}(s)e^{2ik_0 s_{\perp,\mathrm{sup}}(s)h} \right\}, \tag{1}$$

$$\frac{dF_P^{||}}{ds} = \frac{3}{4}\frac{1}{\varepsilon_{\mathrm{sup}}^{1/2}}\mathrm{Re}\left\{ \frac{s}{s_{\perp,\mathrm{sup}}(s)} \left[ 1 + \frac{s_{\perp,\mathrm{sup}}^2(s)}{\varepsilon_{\mathrm{sup}}} + \left( \tilde{r}^{\mathrm{s}}(s) - \frac{s_{\perp,\mathrm{sup}}^2(s)}{\varepsilon_{\mathrm{sup}}} \tilde{r}^{p}(s) \right) e^{2ik_0 s_{\perp,\mathrm{sup}}(s)h} \right] \right\}, \tag{2}$$

$$\frac{dF_P^{ave}}{ds} = \frac{1}{3}\frac{dF_P^{\perp}}{ds} + \frac{2}{3}\frac{dF_P^{||}}{ds}, \tag{3}$$

where $s = k_{||}/k_0$, $s_{\perp,\mathrm{sup}}(s) = k_{\perp,\mathrm{sup}}(s)/k_0 = \ \varepsilon_{\mathrm{sup}} - s^2 \ ^{1/2}$.

The Fresnel reflection coefficients for p- and s-polarized light $\tilde{r}^{\mathrm{p}}$ and $\tilde{r}^{\mathrm{s}}$ were calculated using analogue of Drude formula [4].

## S5. Lasing experiments

The nanorod arrays are spin coated with a 10 mM solution of R101 dye dissolved in PVA at 1000 rpm for 30 s followed by baking at 60 °C for 6 hours, which yields a ~2 μm-thick film. A frequency-doubled Nd:YAG picosecond laser (532 nm, 400 ps pulse width, and 1 Hz repetition rate) is used to pump the samples at 40°, as shown in Figure S4. The pump pulses are focused down to a spot size of ~200 μm in diameter using a 5× objective lens. The emission from the samples is collected with a fiber almost normal to the sample surface, which is fed to a spectrometer (SP-2150i, Princeton Instruments) equipped with a charge-coupled device (CCD).

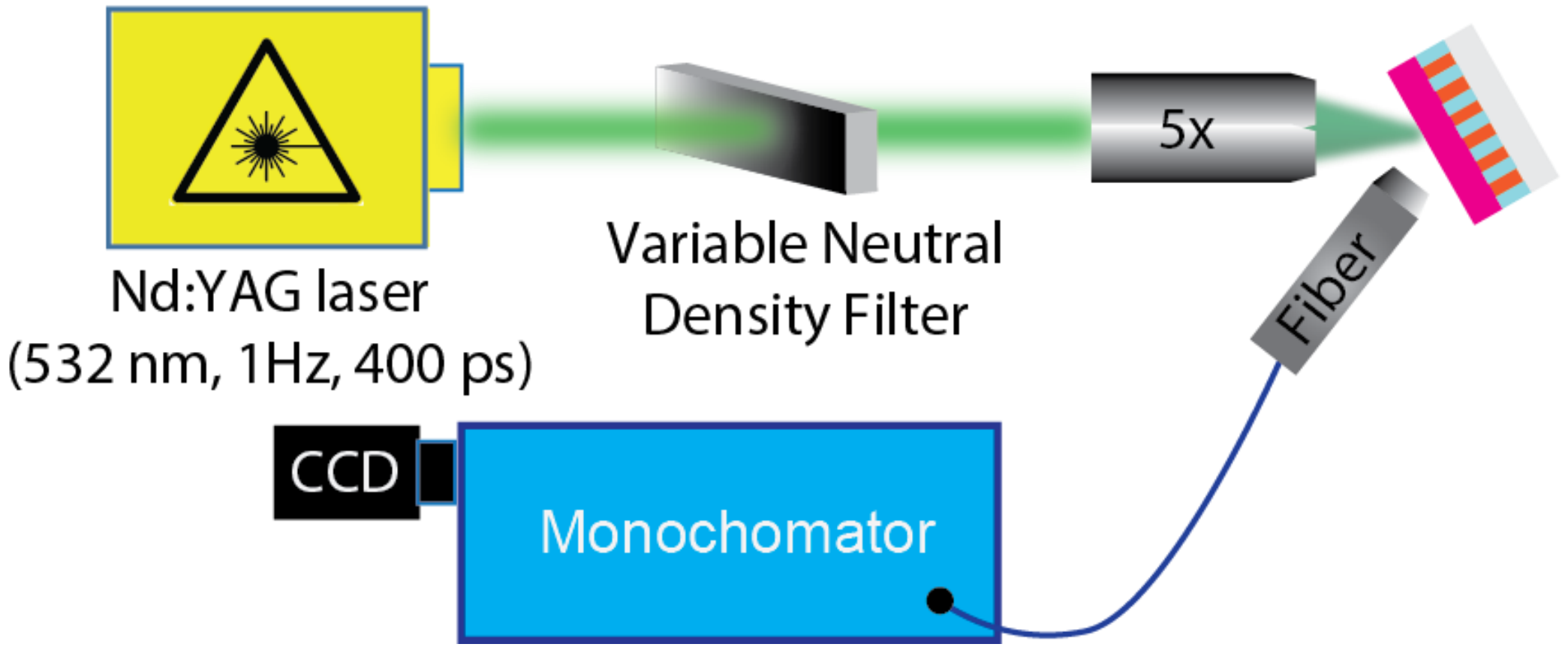

Figure S4. Schematic of setup used to collect lasing emission from nanorod metamaterial samples.

## S6. Spectral narrowing and threshold distribution for nanorod HMM and EMM

One proof of lasing action is a significant reduction in the full width at half maximum (FWHM) of the emission spectra as the pump energy is increased. The FWHM for the emission spectra of nanorod HMM and EMM are shown below in Figure S5. The linewidth of the lasing emission from the nanorod HMM is reduced from ~24 nm to ~5.8 nm as the pump energy is increased to above the threshold. This linewidth is comparable to various other demonstrated lasing systems using dyes as gain media. For the nanorod EMM, we see a reduction of the linewidth from ~24 nm to ~6.18 nm. We see the reduction in FHWM for the HMM at lower pump energy as compared to EMM, hence exhibiting lower threshold.

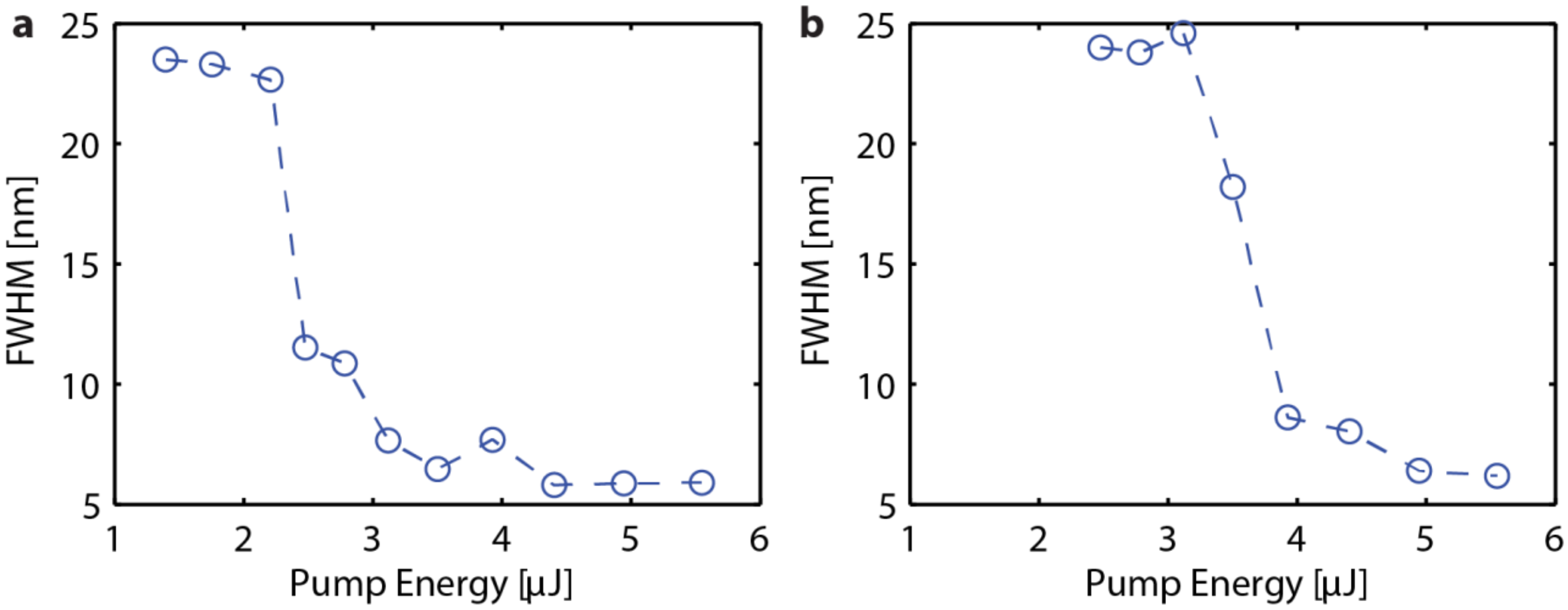

Figure S5. Spectral linewidth of the emission as a function of the pump energy for nanorod HMM (a) and EMM (b) samples, respectively.

## S7. Threshold distribution for nanorod HMM and EMM

We have studied the threshold behavior for nanorod HMM and EMM samples at 5 different points from each sample. The distribution of threshold values for both samples is shown below in Figure S6. We see that on average the HMM gives us a reduction of ~35% in threshold compared to the EMM sample.

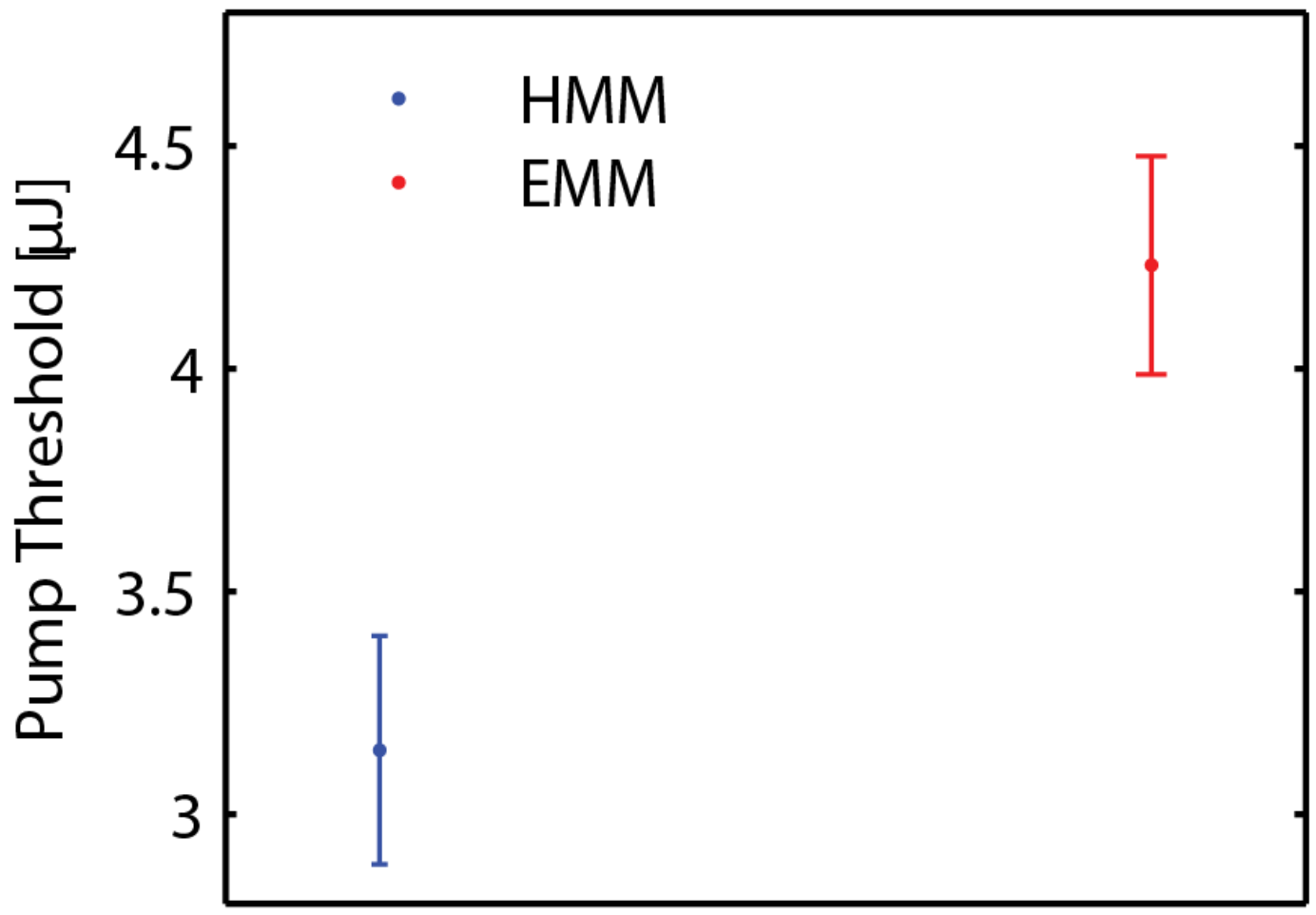

Figure S6. The average pump thresholds of HMM and EMM samples respectively, obtained by measuring five different points from each sample.

## S8. Emission measurements with control samples

We measured the emission from a bare glass substrate and 250 nm-thick gold film deposited on glass, both coated with a 2 μm layer of PVA embedded with R101 where the concentration of R101 is 10 mM relative to PVA. The results are shown in Figure S7 below. Both samples exhibit broadband emission, but show no lasing peaks.

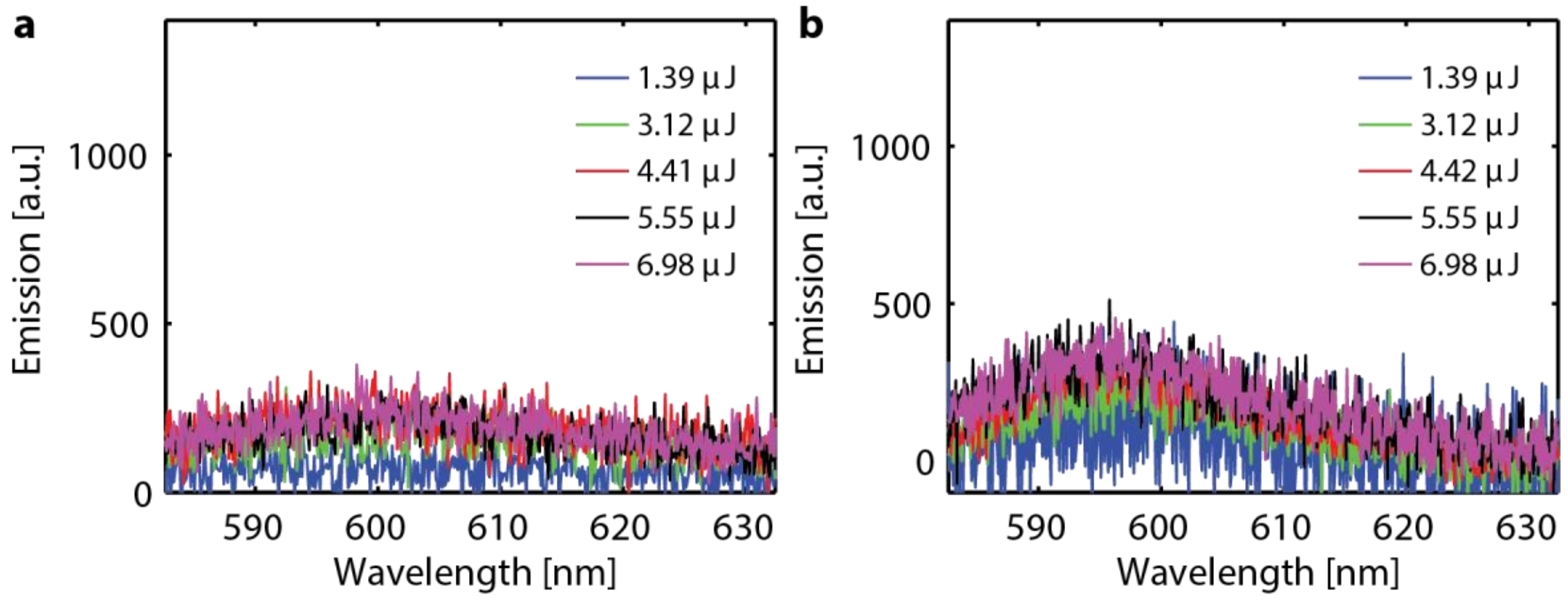

Figure S7. Emission measurements from (a) glass substrate and (b) 250 nm gold film on glass samples coated with a 2μm layer of R101 embedded in PVA.

### S9. Emission from lamellar HMM sample

We measured the emission from a lamellar metal-dielectric stack of alumina and gold layers coated with R101 in PVA. The structure is composed of 10 pairs of a 8.9 nm-thick gold layer and a 16.1 nm-thick alumina layer, which has the same thickness (250 nm) and metal fill ratio (~35%) as the nanorod HMM. The emission measurements are shown in Figure S8a. The plots only shows spontaneous emission and no lasing peaks are observed. We have also quantitatively compared the emission intensity from lamellar and nanorod HMMs, shown in Figure S8b. We can see that the emission from nanorod HMM increases more rapidly than that from lamellar HMM, although the emission intensity from the latter is stronger than the former at low pump energy.

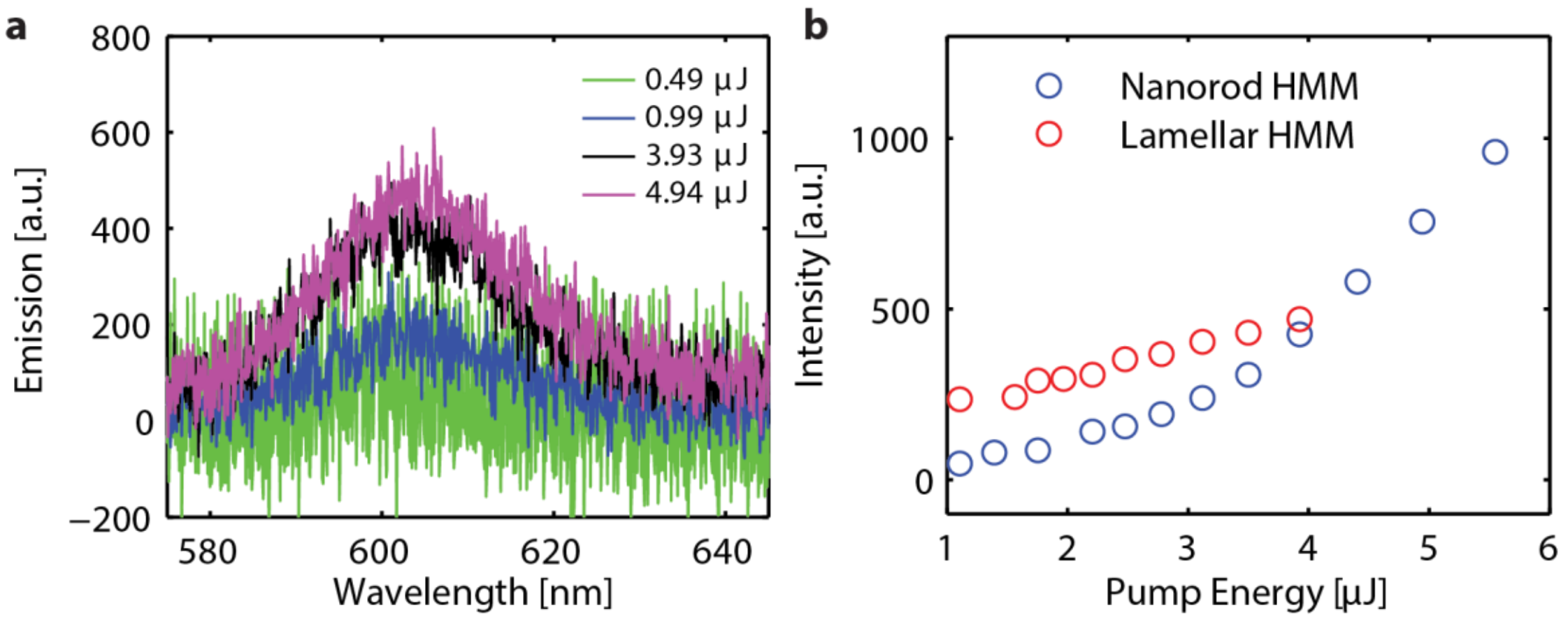

Figure S8. (a) Emission from Lamellar HMM composed on 10 pairs of 8.9 nm gold film and 16.1 nm alumina film, coated with R101 in PVA. Plots show only broad spontaneous emission bands. (b) Comparison of emission intensity from nanorod and lamellar HMMs.

### S10. Local mode and field distributions

The mode and field distributions of gold nanorod metamaterials are calculated with COMSOL Multiphysics 5.2a. The simulated domains are set as reported in Reference 5. The material properties are as provided in section S2.

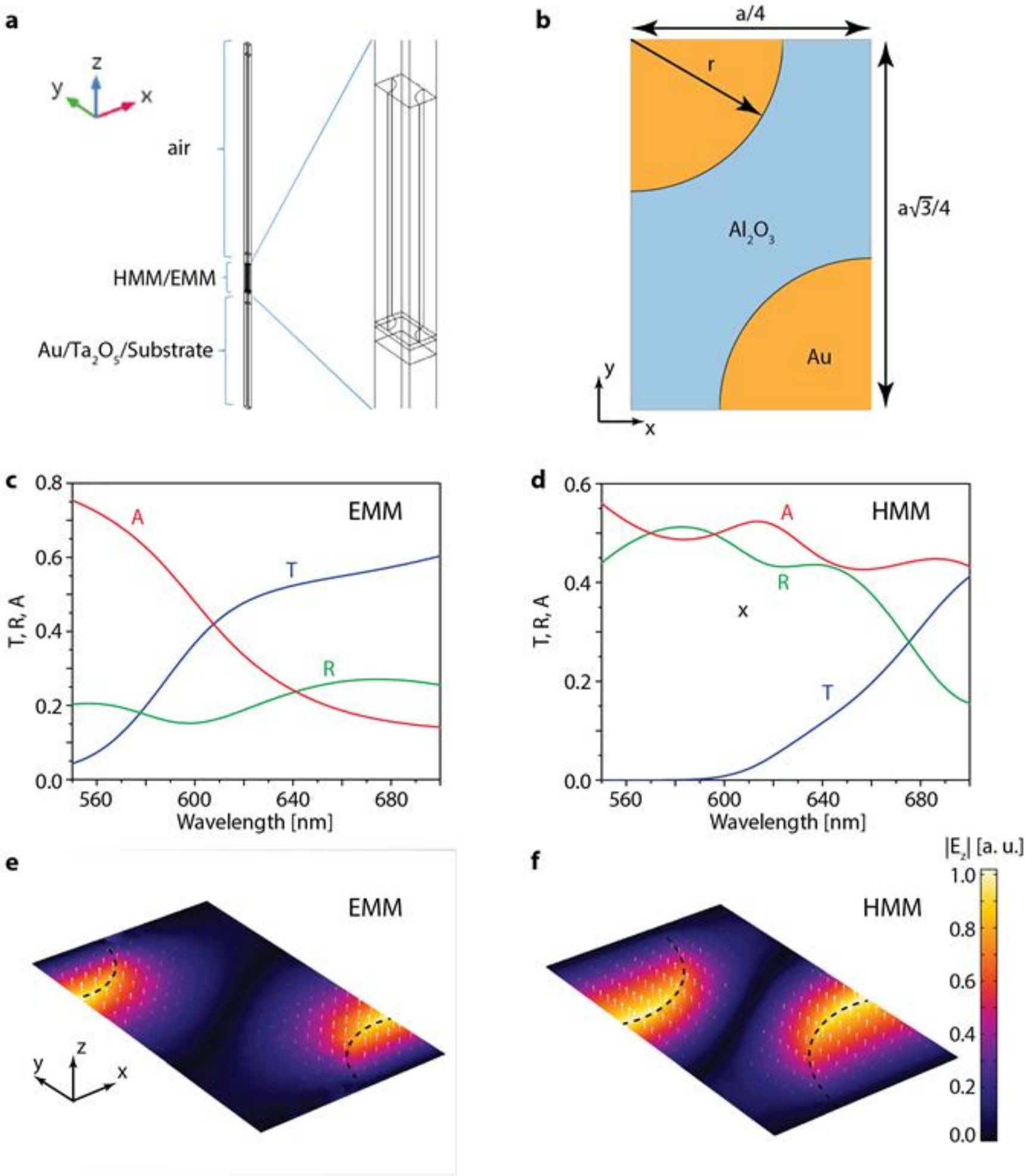

Figure S9. (a) Simulation domain, (b) unit cell of structure, (c,d) calculated optical spectra of EMM (c) and HMM (d): transmission (blue), reflection (green), and absorption (red), (e,f) |Ez| field distribution of the unit cell shown in (b) for EMM (e) and HMM (f). The dashed lines indicate the position of the nanorod. The magnitude of |Ez| field is normalized with respect to the maximum value that is the same for HMM and EMM

## S11 Loss compensation of plasmonic resonance in HMM

In order to estimate whether the losses of the mode at ~606 nm is compensated by the gain, we calculated the optical absorption of a structure having the same dimensions with that shown in Figures S9a, b but the nanorod HMM is covered with a layer of 2 μm-thick R101-PVA. The calculated optical absorption is negative at 606 nm, indicating that the losses are compensated by the gain provided by R101 used in this work.

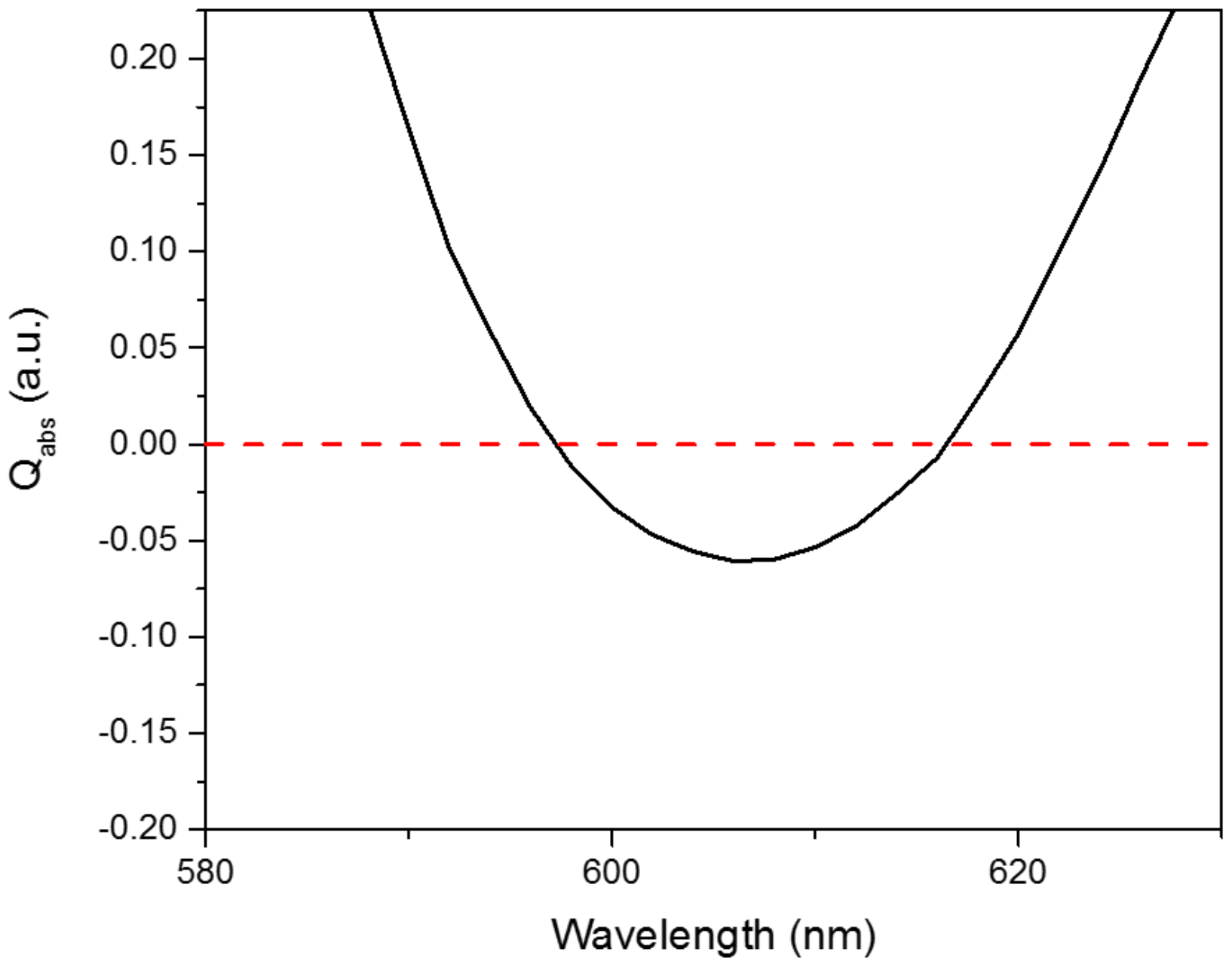

Figure S10. Optical absorption spectrum of the system with gain. In the calculation, the concentration of R101 is 10 mM, corresponding to a number density of $6.02 \times 10^{18}$ $cm^{-3}$. The emission cross section of R101 is set as $2.5 \times 10^{-16}$ $cm^2$.